\documentclass[twocolumn,preprintnumbers,amsmath,amssymb,aps,pra,floatfix%linenumbers
]{revtex4}

\usepackage{units}

\usepackage{isomath}
\usepackage{upgreek} % to have roman greek letter

\usepackage{epsfig, amsmath,amsfonts, amssymb,graphicx,amsthm,color}

\usepackage{mathtools}% For \MoveEqLeft
\usepackage{dcolumn}% Align table columns on decimal point
\usepackage{bm}% bold math
\usepackage{rotating}
\usepackage{amssymb}
\usepackage[latin1]{inputenc}
\usepackage{graphicx}

\usepackage{epstopdf}
\newcolumntype{M}[1]{>{\raggedright}m{#1}}

\begin{document}

\title{Ion and Electron Ghost Imaging}

\author{A. Trimeche\footnote{present address: Universit\'e Paris-Saclay, Institut d'Optique Graduate School, CNRS, Laboratoire Charles Fabry, 91127, Palaiseau, France.}}
\author{C. Lopez\footnote{present address: Universit\'e Paris-Saclay, ENS Paris-Saclay, LUMIN, 91405, Orsay, France}}
\author{D. Comparat}
\author{Y.J. Picard}
\email[Corresponding author: ]{yan.picard@universite-paris-saclay.fr}
\affiliation{Universit\'e Paris-Saclay, CNRS, Laboratoire Aim\'{e} Cotton, 91405, Orsay, France}

\date{\today}

%\pacs{}

\begin{abstract}

In this Letter, we report a demonstration of ion and electron ghost imaging. Two beams of correlated ions and electrons are produced by a photoionization process and accelerated into opposite directions. Using a single time and position sensitive detector for one beam, we can image an object seen by the other beam even when the detector that sees this object has no spatial resolution. The extra information given by this second detector can, therefore, be used to reconstruct the image thanks to the correlation between the ions and the electrons. In our example, a metallic mask placed in front of a time-sensitive detector is used as the object to image. We demonstrated ion and electron ghost imaging using this mask in a transmission mode. These primary results are very promising and open applications especially in ion and electron imaging in surface science and nanophysics. 

\end{abstract}

\maketitle

%%%%%%%%%%%%%%%%%%%%%%%%%%%%%%%%%%%%%%%%%%%%%%%%%%%%%%%%%%%%%%%%%%%%%%%%%%%%
% \section{Introduction }
%%%%%%%%%%%%%%%%%%%%%%%%%%%%%%%%%%%%%%%%%%%%%%%%%%%%%%%%%%%%%%%%%%%%%%%%%%%%

%\section{Introduction}

Ghost imaging also called ``coincidence imaging''  or ``correlated imaging'' is able to
 produce an image of an object even when the detector that sees the object has no spatial resolution. The requirement for this imaging is that a second detector collects a partner particle correlated to the one sent to the object. The extra information given by the second detector can therefore be used to infer the location of the first particle and can thus be used to reconstruct the image.
 Obviously this method is very useful when  spatially resolved detectors are unavailable or when the experimental architecture makes difficult to implement one. It has been demonstrated using (entangled or correlated) photons \cite{strekalov1995observation,pittman1995optical,bennink2002two,gatti2004ghost}, 
  atoms  \cite{khakimov2016ghost} or with
 one photon and one electron \cite{li_electron_2018}.
 
 Here we present this method using correlation between one electron and one ion. We demonstrate both ghost ion imaging using spatially resolved electron detector and 
ghost electron imaging using spatially resolved ion detector.
These methods can then be applied on electron or ion-based imaging systems to reduce image acquisition time or to reduce the sample damage by reducing the amount of particles sent to the sample \cite{spence2017outrunning}. This can be combined with compressed-sensing optimizing  the sparsity of a signal  \cite{katz_compressive_2009} and can thus improve the contrast comparing to conventional image  \cite{morris_imaging_2015}. It will open some spatial resolution capabilities to spectroscopic methods having usually poor spatial resolutions \cite{handbook2019springer} (high resolution electron energy loss spectroscopy (HREELS) being one obvious example). \\

%\section{Experimental apparatus}

%\begin{figure}
%	\centering
%	\includegraphics[width=1.0\linewidth]{Fig1_setup.png}
%	\caption{Scheme of the experimental setup.  Electrons and ions, produced by photoionization of a cesium beam, are extracted in opposite directions by an electric field towards two detectors. The acquisition system handles the information from the detectors and processes the data in a coincidence mode.}
%	\label{fig:experimental-apparatus}
%\end{figure}

The experimental setup, designed to produce beams of correlated ions and electrons, has been described in detail previously \cite{Picard2019realtime} and only a brief description of the parts that are relevant for this work will be given here.
The setup is based on a double time-of-flight (TOF) spectrometer with detectors at opposite ends monitored in coincidence mode. For this experiment, we use cesium (Cs) atoms so as to create ion-electron pairs. Cs effuses from an oven and propagates to the ionization region. Near-threshold Photoionization is performed in a static electric field produced in between two holed electrodes. By using narrowband lasers, in a  three-photon transition process, Doppler selection is performed to reduce drastically the effect of the effusive atomic beam velocity dispersion. After ionization, electron and ion are accelerated by the static electric field in opposite directions toward the detectors.
Since we used the same setup to obtain both ion and electron ghost images and for clarity purposes, the two correlated particles produced by the ionisation will be called particles $A$ and $B$ in the next paragraph, $A$ being the one that sees the object to image, and $B$ the one that will reveal the ghost image of the object.

\begin{figure}[!ht]
	\centering
	\includegraphics[width=1.0\linewidth]{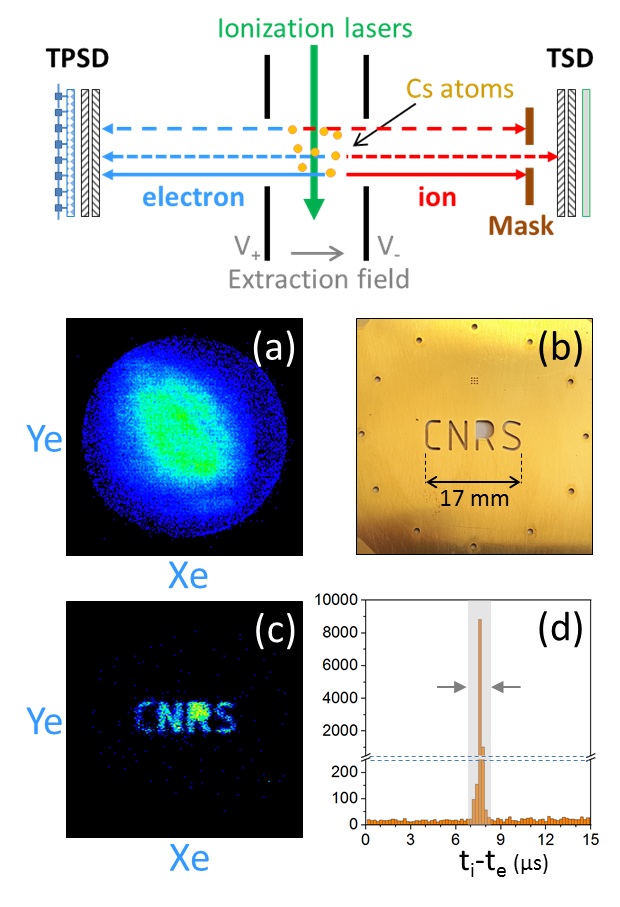}
	\caption{Upper part: Scheme of the experimental setup in the ion ghost imaging configuration. The ion-electron pairs trajectories shown here are schematics and not at scale. (a): Image of the time and position sensitive detector (TPSD) for all the electrons (in blue). (b): Picture of the mask placed in front of the Time Sensitive Detector (TSD). (c): Image of the TPSD only for electrons that are detected in coincidence with their correlated ion. (d): Relative Time of Flight (TOF) histogram. The selection of the event whose relative TOF ($t_i - t_e$) fits into the coincidence window (gray zone in between arrows) gives image (c), whereas the selection of all the events independently of their relative TOF gives image (a).}
	\label{fig:ghosti}
\end{figure}

\begin{figure}[!ht]
	\centering
	\includegraphics[width=1.0\linewidth]{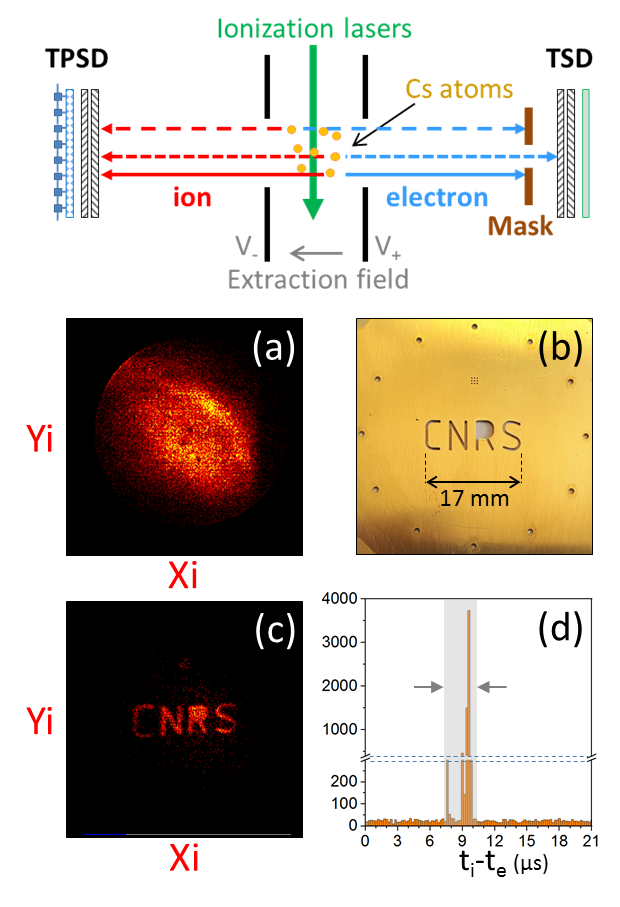}
	\caption{Upper part: Scheme of the experimental setup in the electron ghost imaging configuration. The ion-electron pairs trajectories shown here are schematics and not at scale. (a): Image of the time and position sensitive detector (TPSD) for all the ions (in red). (b): Picture of the mask placed in front of the Time Sensitive Detector (TSD). (c): Image of the TPSD only for ions that are detected in coincidence with their correlated electron. (d): Relative Time of Flight (TOF) histogram. The selection of the event whose relative TOF ($t_i - t_e$) fits into the coincidence window (gray zone in between arrows) gives image (c) whereas the selection of all the events independently of their relative TOF gives image (a).}
	\label{fig:ghoste}
\end{figure}

The correlation between electron and ion allows to infer the position of a particle from its correlated one. On one side of the spectrometer, the mask shown in figures \ref{fig:ghosti} (b) and \ref{fig:ghoste} (b), is placed in front of a time sensitive detector (TSD) composed of a set of micro-channel plates (MCP) and a conductive anode. Thus, a particle $A$ hits this detector only if it passes through the mask. Note that the front face of the TSD and the mask are maintained at the same voltage. The TOF of particle $A$ is measured but its location on the detector can't be determined since no device is set to measure the particle position on that side of the spectrometer.
On the other side of the spectrometer, a time and position sensitive detector (TPSD) composed of a set of $40~{\rm mm}$ open diameter MCP and a delay line device measure the TOF and the position of the correlated particle $B$.
The TOF of particle $A$ and the TOF and position coordinates $X,Y$ of particle $B$ are monitored in coincidence mode by an acquisition system that allows to measure the relative TOF of the correlated particles $A$ and $B$. If this relative TOF fits in a given range around the expected value, the pair is considered as coming from a unique ionization event and is labeled coincident in time. The image obtained from the TPSD can thus be displayed only for the particles $B$ that are measured in coincidence with their correlated particle $A$ to reconstruct the so called ghost image.\\

%\section{Ion Ghost Imaging}

We first start with the configuration called ion ghost imaging (iGI) where all the voltages are set in a way that the electrons are accelerated toward the TPSD and the ions are accelerated toward the mask and the TSD, as shown in the scheme of figure \ref{fig:ghosti}. The mask used in this proof of principle experiment (Fig. \ref{fig:ghosti} (b)) is a $80~{\rm mm}$-diameter brass disk in which we engraved the $0.3~{\rm mm}$ line width ``CNRS'' characters and some holes made for centering purpose. In this case, the ions (in red) will only be detected if they pass through the letters of the mask.  On the other side, the extracted electrons (in blue) fly through the opposite part of the spectrometer and are all detected by the TPSD. Figure \ref{fig:ghosti} (a) shows the raw image of all the electrons detected by the TPSD at the end of an acquisition sequence. Without any treatment, this image can only inform us about the beam spread and its initial conditions.
However, if we display only the electrons that are each in coincidence with an ion using the TOF signals delivered by the TSD and the TPSD, we obtain the image displayed on figure \ref{fig:ghosti} (c). Here we clearly see on this selected TPSD image a ghost image of the mask placed in front of the TSD. This imaging process is based on the coincidence criterion and the position correlation quality.

The coincidence criterion ensure that the detected electron and ion pairs are each coming from the same ionisation event. This is validated by the measurement of the relative TOF of the two charged particles. The electron-ion pairs are created and accelerated in two opposite directions, thus their relative TOF depends mainly on the acceleration voltage and the distances between the extraction zone and the correspondent detectors. As seen on figure \ref{fig:ghosti} (d), the electron-ion relative TOF in this iGI configuration is about $7-8~\mu{\rm s}$. All the events that are outside this relative TOF window correspond to false coincidence, that means signals that are not coming from a Cs ionisation event, such as noise, dark counts, missing counts or different ionized element than Cs. 

The position correlation quality ensure that, for the detected ions that pass through the mask, their corresponding coincident electrons will have a specific spatial distribution. The quality of link between the ion and electron positions depends on the accuracy of the correlation between them. If there is no correlation between the coincident charged particles, the image obtained after filtering the coincident electrons would be also a spot very similar to the one on figure \ref{fig:ghosti} (a). However, if the particle position are correlated, the final position of the electron would depend on the final position of the ion. In this case the image obtained after filtering the coincident electrons would be modulated since the detected ions pass through specific regions of the mask, here the ``CNRS'' letters. This spatial modulation of the coincident electrons depends on the correlation quality.
 The ghost imaging reproduces directly the mask only if the final $(X_i,Y_i)$  ion position and the final $(X_e,Y_e)$ electron position are proportional, that means perfectly thin and linear correlation curves in $X$ and $Y$ coordinates ($X_i$ versus $X_e$ and $Y_i$ versus $Y_e$). This is nearly the case in our system that presents indeed a very thin and linear correlation.
The thinness of the correlations are not limited  by the resolution of the electron TPSD but by the ionization conditions. The optimization of these conditions (Doppler selection, ionization at threshold, proper spatial section of the ionization region, ...) are discussed in details in Ref. \cite{Picard2019realtime}.
The result is that our system has a quite good correlation quality and this allowed us to get the clear images of figures \ref{fig:ghosti} (c).\\

%\section{Electron Ghost Imaging}

We now present the configuration called electron ghost imaging (eGI) where all the voltages are set in a way that the ions are accelerated toward the TPSD and the electrons are accelerated toward the same mask as used before and the TSD, as shown in the scheme of figure \ref{fig:ghoste}. In this case, the electrons (in blue) will only be detected if they pass through the letters of the mask.  On the other side, the extracted ions (in red) fly through the opposite part of the spectrometer and are all detected by the TPSD. Figure \ref{fig:ghoste} (a) shows the raw image of all the ions detected by the TPSD at the end of an acquisition sequence. 
If we now display only the ions that are each in coincidence with an electron using the TOF signals delivered by the TSD and the TPSD, we obtain the image displayed on figure \ref{fig:ghoste} (c). Here we clearly see on this selected TPSD image a ghost image of the mask placed in front of the TSD. As explained previously in the iGI configuration, this imaging process is based on the coincidence criterion and the position correlation quality. For the coincidence criterion, as seen on figure \ref{fig:ghoste} (d), the electron-ion relative TOF in this eGI configuration is about $8-9~\mu{\rm s}$. This value is slightly different from the iGI case because our time-of-flight spectrometer is not symmetric. \\

%\section{Discussion}

The ion ghost-image
(Fig. \ref{fig:ghosti} (c)), as well as the electron one (Fig. \ref{fig:ghoste} (c)), correspond to almost not zoomed and perfect images of the mask (Fig. \ref{fig:ghosti} (b) and \ref{fig:ghoste} (b)). This is fortunate and occurs only because our spectrometer is almost symmetric.  However, we stress that in other conditions, the zoom capability could be a very powerful tool to enhance spatial resolution. It can be used  to see structures that would have been below the spatial resolution of a system. For instance numerous ion and electron setups (time-of-flight mass spectrometry, electron microscopy, Velocity Map Imaging, $\dots$) uses MCP as spatial resolution detector, the resolution being typically $\sim 50\,\mu$m (for segmented anodes, delay lines or phosphor screen) could be enhanced by using zoomed ghost-imaging. Furthermore, providing that the correlation between ion and electron positions stays in a one-to-one correspondence,   a dynamical  or post mathematical treatment of the image can be used to correct the ghost-image   even if some aberrations appear \cite{Picard2019realtime}.   \\

%\section{Conclusion}

 We have performed ghost imaging
using a source of correlated
ions and electrons with
high visibility. Thanks to linear and thin $X$ and $Y$ correlation curves
the ghost images directly represent the physical mask.
In more complex cases of non linear and ideal correlation, for instance due to 
modifications occurring on one trajectory after a zoom or a focusing process,
post treatment of the image is needed to reproduce more precisely the physical mask.

	It is worth mentioning  that the use of the time information from the TPSD and the TSD was useful here only to  unsure the coincidence between the electron and the ion that helps also to reduce background noise.
	However, we would like to stress that the exact same method can be used without any accurate time resolution. A simple Position Sensitive Device (PSD) on one side and a single particle detection capability, one the mask side, is enough to perform the ghost-imaging. The only requirement is that we can, with low ambiguity, correlate the arrival of the particle on the mask side with the correlated particle (that is originating from the same ionization event) on the PSD side. 
Therefore,
our  ghost imaging method can be used in any system presenting a target, for instance using a detector of secondary electrons produced by the particles hitting the sample with only a crude time resolution sufficient enough to be compatible with the repetition rate of the pair production.

Our  ghost imaging  can be more useful for surface science and
nanophysics in a general way
using the second particle to get information that the first one does not provide. For example using energy or time information provided by the other particle (using adequate detectors). 
For instance if the standard analyser is an energy analyser without any spatial resolution such as in electron energy-loss spectroscopy  \cite{egerton_electron_2008}  method, or inversely if the standard analyser is a position sensitive analyser without any energy resolution such as in conventional microscopic 
	Transmission electron microscope (TEM \cite{reimer2013transmission})
	Scanning electron microscope (SEM \cite{goldstein2017scanning}) or Scanning transmission electron or ion microscope (STEM \cite{pennycook2011scanning}, STIM \cite{overley1988energy,breese1992applications}).
	The methods is very general and the information can also be given using not only spatial but for instance using time or energy resolution. 
Furthermore, because the method works for both ions and electrons, it can combine the advantage of both species. As a single example an electron beam has usually a better spatial resolution that an ion one, whereas an ion beam has more chemical and impact properties than electrons. 
 This ion-electron ghost imaging techniques can also be very beneficial for the imagery systems based on electrons or ions since they enables to improve the resolution of the direct imaging systems and to reduce the image acquisition time or to reduce the sample damage by reducing the amount of the probe particles.

These primary results are  promising and can also open the door for  applications especially in quantum physics and fundamental tests like  tests
of EPR entanglement, Bell's inequalities or to improve 
 ion and electron 
interferometry \cite{hasselbach_progress_2009}. \\

%\section{Acknowledgements}

This work was supported by the Fond Unique Interminist\'eriel (IAPP-FUI-22) COLDFIB, the
ANR/DFG HREELM and CEFIPRA No. 5404-1.

\bibliographystyle{h-physrev.bst}
\bibliography{biblio_ghost_imaging}

\end{document}